\documentclass[aps,prd,showpacs,twocolumn,nofootinbib]{revtex4}
\usepackage[dvips]{graphicx}
\usepackage{bm,latexsym,amsmath,amssymb,amsfonts}
\def \D {\text{D}}
\def \c {\text{curl\,}}
\def \dnb {\Delta}

\def \vb {V}

\def \rhog {\rho_{\gamma}}

\begin{document}

\title{
Cosmological magnetic fields from nonlinear effects }

\author{Tsutomu~Kobayashi$^1$, Roy~Maartens$^2$, Tetsuya~Shiromizu$^1$,
Keitaro~Takahashi$^3$}

\affiliation{ $^1$Department of Physics, Tokyo Institute of
Technology,
Tokyo 152-8551, Japan \\
$^2$Institute of Cosmology \& Gravitation, University of
Portsmouth, Portsmouth~PO1~2EG, UK\\
$^3$Department of Physics, Princeton University,
Princeton~NJ08544, USA}


\begin{abstract}

In the standard cosmological model, magnetic fields and vorticity
are generated during the radiation era via second-order density
perturbations. In order to clarify the complicated physics of this
second-order magnetogenesis, we use a covariant approach and
present the electromagneto-dynamical equations in the nonlinear
regime. We use the tight-coupling approximation to analyze Thomson
and Coulomb scattering. At the zero-order limit of exact
tight-coupling, we show that the vorticity is zero and no
magnetogenesis takes place at any nonlinear order. We show that
magnetogenesis also fails at all orders if either protons or
electrons have the same velocity as the radiation, and momentum
transfer is neglected. Then we prove a key no-go result: at
first-order in the tight-coupling approximation, magnetic fields
and vorticity still cannot be generated even via nonlinear
effects. The tight-coupling approximation must be broken at first
order, for the generation of vorticity and magnetic fields, and we
derive a closed set of nonlinear evolution equations that governs
this generation. We estimate that the amplitude of the magnetic field 
at recombination on the horizon scale is $\sim 10^{-27}\,$G.

\end{abstract}


\maketitle

\section{Introduction}

The origin of cosmological magnetic fields is an important problem
in cosmology~\cite{Review}. Many mechanisms for primordial
magnetogenesis (i.e., creation before structure formation) have
been proposed. In order to generate fields on large scales,
inflationary mechanisms are the best candidates, but they require
uncertain modifications to standard physics in order to break the
conformal invariance of Maxwell fields~\cite{Dilaton}.

The generation of cosmological magnetic fields via plasma
interactions during the radiation era, originally suggested by
Harrison~\cite{Harrison}, is based on conventional physics and
does not require any new postulates. The essential ingredients in
this mechanism are nonzero vorticity and Thomson scattering
between photons and charged particles. Because momentum transfer
is more effective between photons and electrons than between
photons and protons due to the mass difference, Thomson scattering
induces differences in the velocity and the distribution of
protons and electrons. These differences induce local electric
currents and net charge density, and the electric field in turn
generates a magnetic field.  This process initiates after
electron-positron annihilation and ends when there are
insufficient free electrons, i.e., it operates over the
temperature range $T_{\rm rec}\lesssim T \lesssim m_e$.

In the standard cosmological model, there is no vorticity at first
order, and the generalized Harrison mechanism is much more
complicated. We can follow the basic argument in a simple
Newtonian formalism. The evolution of the magnetic field is
described by the induction equation,
\begin{equation}
\dot{\vec{B}} = - \vec{\nabla} \times \vec{E}\,.
\end{equation}
Analysis of the momentum transfer in scattering leads to the
generalized Ohm's law,
\begin{equation}
\vec{E} = \eta \vec{J} + \vec{S}\,,
\end{equation}
where $\eta$ is the plasma resistivity and $\vec{S}$ is the
contribution from Thomson scattering. Since $\eta\ll H^{-1}$, we
can neglect this term to obtain
\begin{equation}
\dot{\vec{B}} = - \vec{\nabla} \times \vec{S}\,. \label{Newton}
\end{equation}
As we will see below, $\vec S \sim n \vec v$, where $n$ is the
number density of charged particles and $v$ is the velocity
difference between radiation and charged particles.
Equation~(\ref{Newton}) shows that there are two sources for
magnetogenesis -- vorticity, $\c \vec v$, and the vector product
of density gradient and velocity, $\vec{\nabla}n\times \vec v$.

In the standard cosmology, where perturbations are generated from
inflation, there are no vector modes at first-order, and therefore
the vorticity vanishes at first order. The density-velocity term
is a product of first-order scalar perturbations and therefore
also vanishes at first order. Thus in the standard model a
perturbative analysis of magnetogenesis during the radiation era
must start at second order. (Exceptions can arise if there are
sources of first-order vector perturbations, such as cosmic
strings~\cite{Vachaspati:1991tt}, or fine-tuned anisotropies in
collisionless neutrinos~\cite{Lewis:2004kg}.)

Matarrese et al.~\cite{Riotto} analysed how vorticity and magnetic
fields can be generated from second order cosmological
perturbations. Subsequent work has also used second-order
perturbations~\cite{Gopal:2004ut,Takahashi,Ichiki,Fry,ITS2007},
but has neglected the vorticity and metric vector perturbations,
and focused on the density-velocity terms, i.e., the product of
first-order scalar terms. (For other work on magnetogenesis during
the radiation era, see Refs.~\cite{other}.)
The different approaches lead to estimates in the range~\cite{Riotto,Gopal:2004ut,Fry,ITS2007}:
\begin{equation}\label{est}
B\sim 10^{-24} - 10^{-27}\;\text{G},
\end{equation}
at recombination on 100 Mpc scales.
This is a very weak field, but it provides a seed which is
amplified via the dynamo mechanism. It is possible that the dynamo
amplification can reach the current observed value of about
$10^{-6}$G on galaxy scales~\cite{Dynamo}.

The simplistic Newtonian description given above allows us to
identify the key physical effects, but the real situation is much
more complicated. The second-order perturbative treatments are a
necessary foundation for computing the power spectrum of the
magnetic field. However, it is also useful to adopt a covariant
approach that directly generalises the Newtonian treatment to
cosmology~\cite{Ellis,Roy98}. This allows us to develop a direct
physical understanding of the magnetogenesis process, and also to
deal with the problem in the fully nonlinear regime.

The greatest complexity arises from the dynamics of momentum
transfer via scattering. We use the tight coupling
approximation~\cite{TC}, which is based on the fact that the
scattering time $\tau$ is much less than the cosmic expansion time
$H^{-1}$,
\begin{equation}
 H \tau\ll 1\,,
\end{equation}
so that photons and charged particles are closely bound. In the
limit, i.e., at the zero-order of exact tight-coupling, we have
$\tau=0$ and $v=0$, so that all particles share the same velocity
and behave as a single fluid, and no magnetogenesis takes place.
Beyond the zero-order of tight coupling, there is a nonzero $\tau$
and velocity difference $v$, and the tight coupling approximation
is an expansion in $H\tau $. Note that the tight coupling
approximation is independent of the cosmological perturbative
approximation. Zero-order in tight coupling is not to be confused
with zero-order in cosmological perturbations; the cosmological
variables can be at any nonlinear order, but no magnetogenesis is
possible.

First, we give a basic result: in the limit of exact tight
coupling, and neglecting anisotropic stresses, vorticity vanishes
and magnetic fields cannot be generated at any nonlinear
perturbative order. If the exact tight coupling limit is partially
relaxed by assuming that either the protons or the electrons have
the same velocity as the radiation, but neglecting momentum
transfer, then vorticity and magnetic fields still cannot be
generated at any nonlinear order. Then we derive the evolution
equation for magnetic fields and vorticity beyond the zero-order
of the tight coupling approximation. We show that {\em there is no
magnetogenesis at the first order in the tight-coupling
approximation, and magnetogenesis takes place at second order.}
While it has been suggested that magnetogenesis is possible via
the breaking of the tight coupling limit~\cite{Dolgov}, this is
the first study giving explicitly the condition for such a
mechanism to work.

The rest of the paper is organised as follows. In the next section
we derive the nonlinear equations for magnetic fields and
vorticity in a covariant formalism. In section III, we show that
vorticity cannot be generated even through nonlinear effects in
the tight coupling limit. In section IV we consider the first and
second order of the tight coupling approximation. Finally we
summarise our work in section V.

\section{Cosmological electromagneto-dynamics}

The Faraday tensor can be split into electric and magnetic fields
as measured by a congruence of fundamental observers $u^a$ (with
$u_au^a=-1$):
%
\begin{eqnarray}
F_{ab}=2u_{[a}E_{b]}+\varepsilon_{abc}B^c\,,
\end{eqnarray}
%
where $E_a u^a=B_a u^a=0$. The spatial alternating tensor is
$\varepsilon_{abc}=\eta_{abcd}u^d$, where $\eta_{abcd}$ is the
spacetime alternating tensor, using the convention
$\eta_{0123}=-\sqrt{-g}$. The tensor indices represent an
arbitrary coordinate or tetrad frame; at any event one can choose
local inertial coordinates such that $u^a=(1,\vec 0)$,
$E^0=0=B^0$.

The induced metric in the observer's comoving rest space is
%
\begin{eqnarray}
h_{ab}=g_{ab}+u_a u_b\,,
\end{eqnarray}
%
and it defines a covariant spatial derivative $\D_a$. Generalizing
the Newtonian case, we define kinematical quantities of the $u^a$
congruence via its covariant derivative:
%
\begin{eqnarray}
\nabla_b u_a = \frac{1}{3}\theta
h_{ab}+\sigma_{ab}+\varepsilon_{abc}\omega^c-\dot u_a u_b \,.
\end{eqnarray}
%
Here $\theta = \nabla_a u^a$ is the volume expansion rate,
$\sigma_{ab}=h_a{}^c h_b{}^d \nabla_{(c} u_{d)}$ is the shear,
$\omega_{a}=-{1\over 2}\varepsilon_{abc} \nabla^{b} u^c$ is the
vorticity, and $\dot{u}^a = u^b \nabla_b u^a$ is the acceleration.

Since we are working mainly in the radiation era, it is useful to
choose $u^a$ as the radiation four-velocity in the energy frame,
i.e., with no energy flux:
\begin{equation}\label{radf}
u^a=u_{\gamma}^a\,,~q_{\gamma}^a:=-T_\gamma^a{}_bu_\gamma^b -\rhog
u_\gamma^a =0\,.
\end{equation}
The four-velocities of charged particles, $I=p,e$, are
\begin{equation}\label{pef}
u^a_I=\gamma_I\left( u^a+V_I^a \right), ~u_a V_I^a=0\,,
~\gamma_I=\left( 1 - V_{Ia}V_I^a \right)^{-1/2},
\end{equation}
where we also choose the energy frames, so that $q_I^a:=
-T^a_{Ib}u_I^b-\rho_Iu_I^a=0$.

The Maxwell equations $\nabla_{[a}F_{bc]}=0$ and $\nabla_b
F^{ab}=j^a$ can be split in a 1+3-covariant way (relative to
$u^a$) as~\cite{Ellis}
%
\begin{eqnarray}
\D_a B^a &=& 2E_a \omega^a\,, \\
\D_a E^a &=& -2B_a \omega^a+\mu\,, \\
\dot{B}_a & = & -\frac{2}{3}\theta B_a
+(\sigma_{ab}+\varepsilon_{abc} \omega^c)B^b
\label{bdot} \nonumber \\
& &\quad {}
-\c E_a-\varepsilon_{abc} \dot u^b E^c\,, \label{ind}\\
\dot{E}_a & = & -\frac{2}{3}\theta E_a
+(\sigma_{ab}+\varepsilon_{abc} \omega^c)E^b
\nonumber \\
& & \quad {} +\c B_a +\varepsilon_{abc} \dot u^b B^c -{J}_a\,.
\label{edot}
\end{eqnarray}
%
Here $\mu= -j_au^a$ is the charge density and ${J}_a=h_{ab} j^b$
is the current. For the radiation era plasma with $T\lesssim m_e$,
\begin{eqnarray}
j^a&=&j_p^a+j_e^a\,,~j_I^a=e_In_Iu_I^a\,, \label{j0} \\
\mu&=&e(\gamma_p n_p-\gamma_e n_e)\,,
\\
J^a &=& e \left(\gamma_p n_p V_{p}^a - \gamma_en_e V_{e}^a \right)
, \label{j}
\end{eqnarray}
where $e_I=\pm e$, and $n_{I}$ are the number densities. We can
write
\begin{equation}\label{ni}
n_I=n(1+\Delta_I)\,,
\end{equation}
where $n$ is the density of charged particles in the tight
coupling limit.

The four-current satisfies local charge conservation,
$\nabla_aj^a=0$, which implies
\begin{equation}\label{cc}
\dot{\mu}+\theta\mu=-\D_aJ^a-\dot{u}_aJ^a\,.
\end{equation}
In order to close the Maxwell equations, we need to specify $j^a$,
and this is done via the equations of motion for photons and
charged particles.

In the Maxwell equations we use the covariant curl,
\begin{equation}
\c S_a:= \varepsilon_{abc}\D^bS^c\,,
\end{equation}
and the overdot is the covariant time derivative along the
radiation 4-velocity $u^a$, projected into the rest-space:
\begin{equation}\label{od}
\dot{S}_{a}:=h_a^{~b} u^c\nabla_c S_b\,.
\end{equation}
(This is more convenient than the usual definition, which is not
projected.) At any spacetime event one can choose inertial
coordinates so that $\D_a f=(0,\vec\nabla f)
\dot{S}_{a}=(0,\partial_t \vec S), \c S_a=(0,\vec\nabla\times\vec S)$.
Two important identities are~\cite{Roy98}
\begin{eqnarray}
\D_a\dot f&=& \left(\D_af\right)^{\displaystyle\cdot}
-\dot{f}\dot{u}_a \nonumber
\\&&{} + \left({1\over3}\theta
h_{ab}+\sigma_{ab}-\varepsilon_{abc}\omega^c
\right)\D^bf,\label{id1}\\\c \D_a f &=&-2\dot f \omega_a\,.
\label{id2}
\end{eqnarray}

The vorticity propagation equation is independent of the field
equations, and is given covariantly by~\cite{Roy98}
\begin{equation}
\dot{\omega}_a=-{2\over3}\theta\omega_a+\sigma_{ab}\omega^b
-{1\over2} \c\dot{u}_a\,. \label{vp}
\end{equation}

The conservation law for the electromagnetic energy-momentum
tensor $T_F^{ab} = F^a{}_cF^{bc}-{1\over4}F^{cd}F_{cd}g^{ab}$
follows from the Maxwell equations:
\begin{equation}\label{fc}
\nabla_b T_F^{ab}= - F^{ab}j_b\,.
\end{equation}
Photons and charged particles obey the balance equations
\begin{eqnarray}
\nabla_bT_{\gamma}^{ab}=K_{\gamma}^a\,,~
\nabla_bT_{I}^{ab}=K_{I}^a+ e_In_I F^a{}_bu_{I}^b\,,
\end{eqnarray}
where the $K^a$ four-vectors are the rates of energy-momentum
density transfer to the species. By Eqs.~(\ref{j0}) and
(\ref{fc}), the conservation of the total energy-momentum, $
\nabla_b (T_\gamma^{ab} + \sum_IT_I^{ab} + T_F^{ab}) = 0$,
implies that $K_{\gamma}^a+ K_{p}^a+K_e^a =0$.

The photon energy and momentum balance equations in the general
nonlinear case are~\cite{Roy98}
%
\begin{eqnarray}
\dot \rho_\gamma +{4\over 3}\rho_\gamma\theta &=&
-\sigma_{ab}\pi_{\gamma}^{ab} +U_\gamma\,, \label{ree} \\
{4\over 3}\rho_\gamma\dot{u}^a  +{1\over3}\D^a \rho_\gamma &=&
-\D_b\pi_{\gamma}^{ab}-\dot u_b \pi_{\gamma}^{ab} +M^a_{\gamma}
\,, \label{rme}
\end{eqnarray}
where $\pi_{\gamma}^{ab}$ is the anisotropic stress, with
$\pi_{\gamma}^{ab}u_a=0=\pi_{\gamma}^{ab}h_{ab}$. Here $U_\gamma
=-u_aK^a_{\gamma}$ and $M^a_{\gamma}=h^a{}_b K_{\gamma}^b$ are the
rates of energy and momentum density transfer to photons from
Thomson scattering. From now on, we take $\pi_{\gamma}^{ab}=0$;
the role of photon anisotropic stress in magnetogenesis has been
investigated by Takahashi et al.~\cite{Takahashi,Ichiki,ITS2007}.
For electrons and protons, it is reasonable to neglect pressure
and anisotropic stresses. Then the energy conservation equations
are
\begin{equation}
{u}_{I}^b\nabla_b\rho_I+\theta_I \rho_I=U_I \,, \label{epee}
\end{equation}
where $\rho_I=m_I n_I$ and $U_I = - u_{Ia}K^a_{I}$ is the rate of
energy density transfer due to Thomson and Coulomb scattering. The
momentum balance equations are
\begin{eqnarray}
\rho_{I}{u}_{I}^b\nabla_b u_{I}^a = M^a_{I}
+ e_{I}n_{I} F^a{}_b{u}_{I}^b \,, \label{epme}
\end{eqnarray}
where ${M}_{I}^a=h^a_{Ib} K_{I}^b$ is the rate of momentum
density transfer due to Thomson and Coulomb scattering.

As shown in Maartens et al.~\cite{Roy98}, the Thomson energy
transfer rates, $U_\gamma$ and $U_I$, start from $O(V_I^2)$ and $O(V_I^3)$,
respectively, while the Thomson momentum transfer starts from
linear order in $V_I$. As we will explain later, all $O(V_I^2)$ terms,
except those from Thomson scattering, can be neglected in order
to derive our evolution equation for magnetic fields.
Writing down the necessary terms explicitly, we have
\begin{eqnarray}
U_\gamma &=& \sum_I {\cal C}_{\gamma I} V_I^2
+ O({\cal C}_{\gamma I} V_I^3) \,, \\
U_I &=& O({\cal C}_{\gamma I} V_I^3) \,, \\
M_\gamma^a
&=& - \sum_I {\cal C}_{\gamma I} \left(u^b- u_I^b\right) h^a{}_b
\nonumber \\
&=& \sum_I {\cal C}_{\gamma I} V_I^a + O({\cal C}_{\gamma I} V_I^3) \,,\\
M_I^a
&=& - {\cal C}_{\gamma I} \left(u_I^b-u^b \right)
      h^a_{Ib}-{\cal C}_{IJ}\left(u_I^b-u_J^b \right) h^a_{Ib}
\nonumber \\
&=& - {\cal C}_{\gamma I} (V_I^a + V_I^2 u^a) - {\cal C}_{IJ} (V_I^a-V_J^a)
\nonumber \\
& & + O({\cal C}_{\gamma I} V_I^3)  \,,
\end{eqnarray}
where ${\cal C}_{\gamma I}$, ${\cal C}_{IJ}$ are the Thomson and Coulomb
collision coefficients. The energy and momentum balance equations reduce to
\begin{eqnarray}
\!\! && \dot \rho_\gamma +{4\over 3}\rho_\gamma\theta
= \sum_I {\cal C}_{\gamma I} V_I^2
+ O({\cal C}_{\gamma I} V_I^3) \,, \label{re1} \\
\!\! &&  \dot{n}_I + n_I \left(\theta + \D_a V_I^a + \dot{u}_a V_I^a \right)
+ V_I^a \D_a n_I \nonumber \\
&&\quad\quad{} = O({\cal C}_{\gamma I} V_I^3) + O(V_I^2) \,,
\label{pe1} \\
\!\! && {4\over3}\rho_\gamma\dot{u}^a + {1\over3}\D^a \rho_\gamma
= \sum_I {\cal C}_{\gamma I} V_I^a + O({\cal C}_{\gamma I} V_I^3) \,,
\label{rm1} \\
\!\! && m_I n_I \left(\dot{u}^a+u^b\nabla_bV_I^a+V_I^b \nabla_bu^a\right)
\nonumber \\
&&\quad\quad{} = e_I n_I \left(E^a+F^a{}_b{V}_{I}^b\right)
- {\cal C}_{\gamma I} V_I^a - {\cal C}_{IJ} (V_I^a-V_J^a) \nonumber \\
&&\quad\quad\quad{}
+ O({\cal C}_{\gamma I} V_I^3) \,.
\label{pm1}
\end{eqnarray}
The vorticities of charged particles are related to the radiation
vorticity $\omega^a$ at $O(V_I)$ by~\cite{Roy98}
\begin{eqnarray}\label{wtran}
{\omega}_I^a = \omega^a- {{1\over2}}\c V_I^a + {{1\over2}}
\varepsilon^a{}_{bc} \dot{u}^bV_I^c  +\omega_bV_I^b u^a.
\end{eqnarray}

Taking the tight coupling limit in Eq.~(\ref{pe1}) we recover
number conservation,
\begin{eqnarray}
\dot n+\theta n=0\,,
\end{eqnarray}
and with Eq.~(\ref{re1}), this leads to
$u^a\nabla_a[\ln(n/\rho_{\gamma}^{3/4})]=0$. We define the entropy
\begin{eqnarray}
s_a:=\D_a\left[\ln\left(\frac{n}{\rho_{\gamma}^{3/4}}\right)\right].
\end{eqnarray}
Using the identity~(\ref{id1}), we arrive at
\begin{eqnarray}
\dot s_a+\frac{1}{3}\theta
s_a+(\sigma_{ba}+\varepsilon_{bac}\omega^c)s^b=0\,.
\label{entropy-per}
\end{eqnarray}
In what follows we assume the adiabatic condition $s_a=0$, i.e.,
\begin{eqnarray}
\frac{\D_an}{n}=\frac{3}{4}\frac{\D_a\rho_{\gamma}}{\rho_{\gamma}}\,,
\label{adiabatic}
\end{eqnarray}
which is consistent with Eq.~(\ref{entropy-per}).

\section{Tight coupling limit}

In the exact tight coupling limit, the velocities of photons,
protons and electrons are equal, $V_{I}^a=0$, and the momentum
transfer terms vanish. Intuitively we expect that vorticity and
magnetic fields cannot be generated from zero. In fact, we can
prove a stronger result, that vorticity is zero at all nonlinear
orders. Equation~(\ref{pm1}) reduces to $m_{I}\dot{u}^a= e_I
E^a\,,$ which implies $E_a=0= \dot{u}_a \,.$ Then the photon
momentum balance equation~(\ref{rm1}) reduces to
$\D_a\rho_\gamma=0$, and since $\dot\rho_\gamma \neq0$ by
Eq.~(\ref{re1}), the identity~(\ref{id2}) implies
\begin{equation}\label{v1}
\omega_a=0\,.
\end{equation}
The induction equation~(\ref{ind}) reduces to
%
\begin{eqnarray}\label{i1}
\dot B_a=-\frac{2}{3}\theta B_a+\sigma_{ab}B^b\,,
\end{eqnarray}
%
so there is no source term and no magnetogenesis.
Equations~(\ref{v1}) and (\ref{i1}) hold in the fully nonlinear
regime.

Next we consider what happens if exact tight coupling is weakened
by neglecting scattering terms, and neglecting the velocity
difference between protons and photons, i.e., $V_p^a=0$, but
allowing $V_e^a\neq0$. This is effectively the assumption made in
Ref.~\cite{Riotto}, and here we reconsider the problem in the
covariant formalism. The proton equation of motion~(\ref{pm1})
becomes $m_p\dot{u}^a=eE^a$, and taking the curl gives
\begin{equation}
 \c \dot{u}_a={e\over m_p}\c E_a\,.
\end{equation}
The curl of the photon momentum equation~(\ref{rm1}), using the
photon energy equation~(\ref{re1}) and the identity~(\ref{id2}),
gives
\begin{eqnarray}
\c \dot {u}_a= -\frac{2}{3}\theta \omega_a \,. \label{dotu}
\end{eqnarray}
Using these equations, the induction equation~(\ref{bdot}) and the
vorticity propagation equation~(\ref{vp}) become
%
\begin{eqnarray}
\dot{B}_a &=& -\frac{2}{3}\theta B_a
+(\sigma_{ab}+\varepsilon_{abc} \omega^c)B^b
 +{2m_p\over 3e}\theta\, \omega_a \,,
\label{bdot3}\\ \dot{ \omega}_a &=& -\frac{1}{3}\theta \omega_a
+\sigma_{ab} \omega^b\,.
\end{eqnarray}
%
Thus vorticity and the magnetic field are {\em conserved} and no
magnetogenesis is possible. This gives our first no-go result: \\
{\it If (i)~the scattering terms are neglected and the
proton-photon velocity difference is neglected, (ii)~anisotropic
stress is neglected, (iii)~the initial magnetic field and
vorticity are zero, and (iv)~the energy-momentum conservation
equations hold, then vorticity and magnetic fields cannot be
generated, at any perturbative order.}

The same result holds if we assume the electron velocity equals
the radiation velocity but instead $V_p^a\neq 0$. It is not clear
how the results in Ref.~\cite{Riotto} relate to our no-go result.

\section{Tight coupling approximation}

Nonlinear magnetogenesis is ruled out in the tight coupling limit
of zero collision time, $\tau=0$. Beyond the zero-order of tight
coupling, there is a nonzero $\tau$ and a nonzero velocity
difference $v$, which is governed by the momentum balance
equations~(\ref{rm1}) and (\ref{pm1}). Schematically, these are of
the form
\begin{equation}
\dot v = {v \over \tau}+A\,,
\end{equation}
where $A$ represents terms other than scattering terms. Since
$\dot v\sim Hv$, we have $\dot v \ll v/\tau$. We expand in terms
of the tight coupling parameter $\tau H$:
\begin{equation}
v=v_{(1)}+v_{(2)}+\cdots,~ A=A_{(0)}+A_{(1)}+\cdots,
\end{equation}
and we use TCA($n$) to denote $n$-th order in the tight coupling
approximation. Then
\begin{eqnarray}
\mbox{TCA(1):}~~&& 0={v_{(1)} \over \tau}+A_{(0)}\,, \label{tca1}\\
\mbox{TCA(2):}~~&& \dot{v}_{(1)}= {v_{(2)} \over \tau}+A_{(1)}\,.
\label{tca2}
\end{eqnarray}

The TCA is complicated by the presence of Coulomb scattering, so
that strictly we need to perform TCA expansions in both Thomson
and Coulomb small parameters. However, as we will argue, it is
reasonable to neglect the Coulomb collision time, i.e., to assume
tight coupling of protons and electrons.

The collision coefficients in Eqs.~(\ref{rm1}) and (\ref{pm1}) are
%
\begin{eqnarray}
&& {\cal C}_{\gamma e} = {4\over3} \sigma_T \rho_\gamma n_e \,, \\
&& {\cal C}_{\gamma p} =  \beta^2 {n_p \over n_e}{\cal C}_{\gamma e}\,,
\\&& {\cal C}_{pe} = e^2 n_en_p \eta \,,
\end{eqnarray}
%
where $\sigma_T$ is the Thomson cross-section and
\begin{eqnarray}
\beta:=\frac{m_e}{m_p}.
\end{eqnarray}
Here $\eta$ is the resistivity of the cosmic plasma,
%
\begin{eqnarray}
\eta = \frac{4 \pi e^{2}}{m_e} \left( \frac{m_e}{T} \right)^{3/2}
\ln{\Lambda} \sim 10^{-13} \left( \frac{T}{\rm eV} \right)^{-3/2}
{\rm s},
\end{eqnarray}
%
where $\ln{\Lambda}$ is the Coulomb logarithm, $\Lambda \sim
T^{3/2}e^{-3}n^{-1/2}$ and $n\sim 10^{-10}T^3$. The ${\cal C}$'s
define key timescales, together with the Hubble timescale:
%
\begin{eqnarray}
& & \tau_{\gamma e} := \frac{m_e n_e}{{\cal C}_{\gamma e}}
\sim 10^{5} \left( \frac{T}{\rm eV} \right)^{-4} \,\text{s}, \\
& & \tau_{\gamma p} := \frac{m_p n_p}{{\cal C}_{\gamma p}} =
\beta^{-3} \tau_{\gamma e}
\sim 10^{15} \left( \frac{T}{\rm eV} \right)^{-4} \,\text{s}, \\
& & \tau_{ep} := \frac{m_e n_e}{{\cal C}_{pe}}
\sim 10^{-4} \left( \frac{T}{\rm eV} \right)^{-3/2} \,\text{s}, \\
& & H^{-1} \sim 10^{12} \left( \frac{T}{\rm eV} \right)^{-2}
\,\text{s}.
\end{eqnarray}
%
Thus
%
\begin{eqnarray}
& & H \tau_{\gamma p}
\sim 10^3 \left( \frac{T}{\rm eV} \right)^{-2}, \\
& & H \tau_{\gamma e}
\sim 10^{-7} \left( \frac{T}{\rm eV} \right)^{-2}, \label{ratio-T} \\
& & H \tau_{ep} \sim 10^{-16} \left( \frac{T}{{\rm eV}}
\right)^{1/2}. \label{ratio-C}
\end{eqnarray}
%
As one can see, Thomson scattering between photons and electrons
and Coulomb scattering between protons and electrons are very
effective on cosmological timescales, so that they are tightly
coupled before recombination. Although Thomson scattering between
photons and protons is less effective at low temperatures, protons
also closely follow photons through their Coulomb coupling to
electrons.

From Eqs. (\ref{ratio-T}) and (\ref{ratio-C}), we see that,
%
\begin{equation}
\frac{\tau_{\gamma e}}{\tau_{ep}} \sim 10^9 \left( \frac{T}{\rm
eV} \right)^{-5/2}.
\end{equation}
%
Therefore, at lower temperatures, $T \lesssim 1 ~ {\rm keV}$,
Coulomb scattering is more effective than Thomson scattering, so
that protons and electrons are more tightly coupled than photons
and charged particles are. This suggests that we can safely
neglect Coulomb scattering, i.e., take $\tau_{ep}=0= V_p^a-V_e^a=\Delta_p-\Delta_e $.
We define the centre of
mass velocity $V^a$ and number density deviation $\Delta$:
\begin{eqnarray}
(m_pn_p+m_en_e)V^a &=& m_pn_pV_p^a +m_en_e V_e^a\,,\\
(m_p+m_e)\Delta &=& m_p\Delta_p +m_e\Delta_e\,.
\end{eqnarray}
Then the peculiar velocities decompose as
\begin{equation}\label{vv}
V_I^a=V^a+ v_I^a \approx V^a\,,
\end{equation}
where $v_I^a$ are the deviations of proton and electron velocity
from their centre of mass velocity, with
$m_pn_pv_p^a+m_en_ev_e^a=0$. The number density deviations
decompose as
\begin{equation}\label{dd}
\Delta_I= \Delta + \delta_I \approx \Delta\,,
\end{equation}
where $\delta_I$ are the deviations of proton and electron number
density from their centre of mass density, with
$m_p\delta_p+m_e\delta_e=0$.
The approximations in Eqs.~(\ref{vv}) and (\ref{dd}) are based on
neglecting terms of order $\tau_{ep}$:
%
\begin{eqnarray}
H \tau_{\gamma e} \sim |\Delta| \sim |V^a|  \gg H \tau_{ep} \sim
|\delta_I| \sim |v_I^a|\,.
\end{eqnarray}
%

In fact, protons and electrons are coupled not only by Coulomb
scattering but also by the electric field, so that  $|\delta_I| ,
|v_I^a|$ are further suppressed by a factor~\cite{TIS2007},
%
\begin{equation}
H \eta \sim 10^{-27} \left( \frac{T}{\rm eV} \right)^{1/2}.
\end{equation}
Thus we can safely apply the approximation at any temperatures we
consider here, $m_e \gtrsim T \gtrsim T_{\rm rec}$, and we can
solve equations perturbatively with respect to the small parameter
$H \tau_{\gamma e}$.

In Eqs.~({\ref{pe1})--(\ref{pm1}), we can set $V_I^a=V^a$. We can see
from Eqs. (\ref{tca1}) and (\ref{tca2}) that, upto TCA(2) we consider here,
$O({\cal C}_{\gamma I} V_I^3)$ and $O(V_I^2)$ terms can be neglected and
the momentum equations become,
\begin{eqnarray}
\!\! && {4\over 3}\rho_\gamma\dot{u}^a + {1\over3} \D^a \rho_\gamma
= {(\beta^3 m_pn_p +m_en_e) \over \tau} V^a \,, \label{rm2} \\
\!\! && m_{I} \left(\dot{u}^a + u^b \nabla_b V^a + V^b \nabla_b u^a \right)
\nonumber\\
&&\quad{}\quad =  e_{I} \left(E^a + F^a{}_b{V}^b\right)-
{m_I \over \tau_{I}}\,V^a \,, \label{pm2}
\end{eqnarray}
where we have defined
\begin{equation}
\tau:=\tau_{\gamma e}\,,~\tau_I:=\tau_{\gamma
I}=(\beta^{-3}\tau,\tau) \,.
\end{equation}
On the other hand, as we will see in section \ref{subsection:tca2},
we need only TCA(1) for the charged particle conservation equation
(\ref{pe1}) to derive the evolution equation:
\begin{equation}
\dot\Delta+\D_aV^a+\dot{u}_aV^a+V^a \frac{\D_a n}{n} =0\,.
\label{ndc}
\end{equation}

\subsection{First order -- TCA(1)}

At first-order TCA, we follow Eq.~(\ref{tca1}) and keep only the
first-order $V^a$ and the zero-order of other terms.

By summing the photon and charged particle equations~(\ref{rm2})
and (\ref{pm2}), we obtain
\begin{equation}
\dot{u}^a = - \frac{1}{4 \rhog(1+R)}\, \D^a \rhog\,,\label{geom2}
\end{equation}
where
\begin{equation}
R:={3\rho_b \over 4\rhog}\,,~\rho_b:= (m_p + m_e) n\,.
\end{equation}
The charged particle equations~(\ref{pm2}) imply
%
\begin{eqnarray}
0 = (1 + \beta) \frac{e}{m_e} \,E^a +  {(1 - \beta^3)\over \tau}
\, \vb^a_{(1)}\,. \label{peeom}
\end{eqnarray}
The photon and charged particle equations~(\ref{rm2}) and
(\ref{pm2}) also lead to
%
\begin{eqnarray}
0 = \frac{1}{4 \rhog}\, \D^a \rhog - \left(\frac{1 + \beta^2}{1 +
\beta}\right) \frac{\beta(1+R)}{\tau}\, \vb^a_{(1)}\,.
\label{beom}
\end{eqnarray}

It follows that
%
\begin{eqnarray}
E^a &=& - \frac{m_e}{e}\frac{(1 - \beta^3)}{(1 + \beta)}
\frac{1}{\tau}\, \vb^a_{(1)}
\nonumber \\
&=& - \frac{m_p}{e}\frac{(1 - \beta^3)}{(1 + \beta^2)} \frac{1}{4
\rhog(1+R)}\, \D^a \rhog\,. \label{Ohm3}
\end{eqnarray}
%
Using Eqs.~(\ref{geom2}) and (\ref{Ohm3}) in the induction
equation~(\ref{ind}), we find the evolution equation for the
magnetic field at TCA(1):
%
\begin{eqnarray}
\dot{B}_a & = & -\frac{2}{3}\theta B_a
+(\sigma_{ab}+\varepsilon_{abc} \omega^c)B^b \nonumber\\ &&~
-\left[\frac{m_p}{e} \frac{(1 - \beta^3)}{(1 + \beta^2)}
\frac{\dot{\rhog}}{2 \rhog(1+R)}\right] \omega_a\,,
\end{eqnarray}
%
where we used the adiabatic condition~(\ref{adiabatic}). The
magnetic field is sourced by the vorticity of photons. However,
from Eqs.~(\ref{adiabatic}) and (\ref{geom2}), we see that $\c
\dot{u}^a = [\dot \rho_\gamma/2\rho_\gamma (1+R)]\omega^a$, and
the vorticity propagation equation~(\ref{vp}) becomes
\begin{equation}
\dot{\omega}_a=\left[ \frac{1+2R}{4(1+R)}\frac{\dot
\rho_\gamma}{\rho_\gamma}\right] \omega_a+\sigma_{ab}\omega^b \,.
\end{equation}
This shows that there is no source term for the vorticity, and we
have our second no-go result:\\ {\em No generation of vorticity or
magnetic fields is possible at first order in the tight coupling
approximation.}

Note that at TCA(1), terms of the form $\vec\nabla n\times\vec v$
do not arise. These terms come in at TCA(2). Magnetogenesis via
the $\vec\nabla \times\vec v$ vorticity term is not possible
without breaking the tight coupling approximation at first order;
it is not clear how the results of Ref.~\cite{Riotto} conform to
this no-go result.

\subsection{Second order -- TCA(2) \label{subsection:tca2}}

Now we proceed to the second-order TCA version of the nonlinear
evolution equations for the magnetic field and vorticity.

The charged particle equations~(\ref{pm2}) imply
\begin{equation}
E^a + F^a{}_bV^b = -  \frac{m_e}{e}\frac{(1 - \beta^3)}{(1 +
\beta)} \frac{1}{\tau} V^a\,. \label{Ohm_full}
\end{equation}
Substituting into the induction equation~(\ref{ind}), we have
\begin{eqnarray}
\!\!\!\!\! && \dot{B}_a = -\frac{2}{3}\theta B_a
+(\sigma_{ab}+\varepsilon_{abc} \omega^c)B^b \nonumber\\ && \quad
{} +\frac{m_e}{e} \frac{(1 - \beta^3)}{(1 + \beta)} \frac{1}{\tau}
\left[ \c V^a - \frac{3}{4} \varepsilon^{abc}V_b \frac{\D_c
\rhog}{\rhog}  \right]\!\!,\label{ind2}
\end{eqnarray}
where $F^a{}_bV^b$ in Eq.~(\ref{Ohm_full}) was dropped because it
is always negligible compared to $\dot{B}^a$.

We need to solve for $V^a$ at TCA(2). The photon and charged
particle equations~(\ref{rm2}) and (\ref{pm2}) give
\begin{eqnarray}
\!\!\! && u^b\nabla_b {V}^a + V^b \nabla_b u^a  - \frac{1}{4}
\frac{\D^a \rhog}{\rhog}
\nonumber \\
&& ~~ = - \frac{\beta(1 + \beta^2)}{(1 + \beta)}{( 1 + R + R \dnb
)\over \tau} \,V^a \,.
\end{eqnarray}
This can be split into TCA(1) and TCA(2) equations:
\begin{eqnarray}
\!\!\!\!\!\!  && - \frac{1}{4} \frac{\D^a \rhog}{\rhog} = -
\frac{\beta (1 + \beta^2)}{(1 + \beta)}{( 1 + R)\over \tau}\,
V^a_{(1)}\,,
\label{1st} \\
&& u^b\nabla_b V_{(1)}^a + V^b_{(1)} \nabla_b u^a \nonumber \\
&& ~ = - \frac{\beta(1 + \beta^2)}{(1 + \beta)}{1\over \tau}\left[
( 1 + R ) V^a_{(2)} + R \dnb_{(1)} V^a_{(1)} \right]\!\!,
\label{2nd}
\end{eqnarray}
where the first equation is equivalent to Eq.~(\ref{beom}). We can
solve these equations for $V^a$:
\begin{eqnarray}
\!\!\! && V^a_{(1)}
= \frac{1}{4(1+R)} \frac{(1 + \beta)}{\beta (1 + \beta^2)}
\tau \frac{\D^a \rhog}{\rhog}\,, \label{1stV} \\
&& V^a_{(2)}
= - \frac{R}{4(1+R)^2} \frac{(1 + \beta)}{\beta (1 + \beta^2)}
\tau \dnb_{(1)} \frac{\D^a \rhog}{\rhog}
\nonumber \\
& &~ {}- \frac{1}{(1+R)} \frac{(1 + \beta)}{\beta (1 + \beta^2)}
\tau \left[ u^b\nabla_b{V}^a_{(1)} + V^b_{(1)} \nabla_b u^a
\right]\!\!. \label{2ndV}
\end{eqnarray}
The last term on the right of Eq.~(\ref{2ndV}) is determined in
terms of $\D^a\rhog$ from Eq.~(\ref{1stV}), and the vorticity
occurs explicitly since $V^b_{(1)} \nabla_b u^a\propto
\D^b\rhog[{1\over3} \theta
h^a{}_b+\sigma^a{}_b+\varepsilon^a{}_{bc}\omega^c]$.

Now we can compute the crucial term in Eq.~(\ref{ind2}):
\begin{eqnarray}
\!\!\! & & \c V^a -\frac{3}{4}\varepsilon^{abc}V_b
\frac{\D_c \rho_\gamma}{\rho_\gamma}
\nonumber \\
& &~~ = -\frac{1}{2(1+R)}\frac{(1+\beta)}{\beta (1+\beta^2)} \tau
\frac{\dot \rho_\gamma}{\rho_\gamma}\, \omega^a \nonumber \\
& &~~~~{} -\frac{R}{4(1+R)^2} \frac{(1+\beta)}{\beta (1+\beta^2)}
\tau \varepsilon^{abc} \D_b \Delta_{(1)} \frac{\D_c
\rho_\gamma}{\rho_\gamma}.
\end{eqnarray}

Finally the evolution equation~(\ref{ind2}) for the magnetic field
can be written, up to TCA(2), as
\begin{eqnarray}
\dot{B}^a
&=& -\frac{2}{3}\theta B^a+\sigma^a{}_{b}B^b \nonumber \\
& & {} - \left[\frac{m_p}{e}\frac{R}{4(1+R)^2}
\frac{(1-\beta^3)}{(1+\beta^2)}\right]
\varepsilon^{abc} \D_b \dnb_{(1)} \frac{\D_c \rhog}{\rhog} \nonumber \\
& &{} - \left[\frac{m_p}{e}\frac{1}{2(1+R)}
\frac{(1-\beta^3)}{(1+\beta^2)} \frac{\dot{\rhog}}{\rhog} \right]
\omega^a\,. \label{2nd-B}
\end{eqnarray}
The vorticity evolution equation~(\ref{vp}) can be rewritten,
using Eqs.~(\ref{rm2}), (\ref{1stV}) and (\ref{2ndV}), as
\begin{eqnarray}
\dot{\omega}^a &=& \left[\frac{(1+2R)}{4(1+R)} \frac{\dot
\rho_\gamma}{\rho_\gamma}\right]\omega^a +\sigma^a_{~b}\omega^b
\nonumber \\
& &{}  -\left[ \frac{R}{8(1+R)^2}\right] \varepsilon^{abc} \D_b
\dnb_{(1)} \frac{\D_c \rhog}{\rhog}\,. \label{2nd-omega}
\end{eqnarray}
The evolution of the baryonic number density deviation is governed
by Eq.~(\ref{ndc}), which becomes, up to TCA(1),
\begin{equation}
\dot{\dnb}_{(1)} =
- {3\over4} V^a_{(1)} {\D_a\rhog \over \rhog}
- \D_a V^a_{(1)} - \dot{u}_a V^a_{(1)}, \label{ndc2}
\end{equation}
where $V^a_{(1)}$ is given by Eq.~(\ref{1stV}).

Equations (\ref{2nd-B}), (\ref{2nd-omega}) and (\ref{ndc2}), for
given $\D^a \rhog/\rhog$, form a complete set of equations which
describe the evolution of $\dnb_{(1)}$, $B^a$ and $\omega^a$.
We can see that {\em both magnetic field and vorticity are generated
at the second order in the tight coupling approximation}.

\section{Summary}

In this paper we have derived the evolution equations for
cosmological magnetic fields and vorticity using the 1+3-covariant
formalism. The covariant approach allows us to construct a set of
equations describing the {\em fully nonlinear} evolution of cosmic
inhomogeneities. We have performed a tight coupling expansion for
Thomson and Coulomb interactions to make the key physical
processes transparent. It should be enphasized that we have not expanded
inhomogeneous quantities
with respect to cosmological perturbations, and therefore
our results are valid at any order in cosmological perturbations.
Thus, the present analysis is complementary to previous studies
based on cosmological perturbation theory.

Our first no-go result is that magnetic fields and vorticity
cannot be generated in the tight coupling limit or its weak
extension without anisotropic stresses. Then we have considered
leading and next-to-leading order effects in the tight coupling
approximation. We have found that magnetic fields and vorticity
are not generated at first order in the tight coupling
approximation [TCA(1)]. The second order tight coupling
approximation [TCA(2)] is necessary for generating both of them,
and we have derived a closed set of nonlinear evolution equations
at TCA(2). It is worth noting that we have not invoked the
Einstein equations, so that our result does not rely on any
specific theory of gravity.

The magnitude of the generated magnetic field can be roughly
estimated as follows. From Eqs. (\ref{1stV}) and (\ref{ndc2}),
we have $\dnb_{(1)} \sim (\tau k^2 \delta)/(\beta H a^2)$ where
$\delta \sim 10^{-5}$ is the density perturbation, $k$ is
the wave number and $a$ is the scale factor. Using Eqs. (\ref{2nd-B})
and (\ref{2nd-omega}), we see that the contributions to $B^a$
from the vorticity term and the gradient term are of the same order
of magnitude, and we obtain
\begin{equation}
\langle |B| \rangle \sim \sqrt{B_a B^a}
\sim \frac{m_p R \tau}{e \beta H^2} \left( \frac{k}{a} \right)^4
       \delta^2
\sim 10^{-27} {\rm G},
\end{equation}
where we evaluated the amplitude at the horizon scale at
recombination, $H \sim k \sim 1/(100 {\rm Mpc})$ and $a \sim 10^{-3}$.
This is in the range of previous estimates
\cite{Riotto,Gopal:2004ut,Fry,ITS2007}, as in Eq. (\ref{est}). 

The anisotropic stress of photons is neglected in the present
analysis. However, as reported by Ichiki et
al.~\cite{Ichiki,ITS2007}, this is important for magnetogenesis on
small scales ($\lesssim$ 1 Mpc) and in the earlier universe. It is
expected that the anisotropic stress is important also for the
generation of vorticity on the same scales and in the same era. We
will discuss the effect of the anisotropic stresses in future
work.

\[ \]{\bf Acknowledgements:}

TK and TS thank Akio Hosoya and Toshio Terasawa for their useful
comments. RM thanks Chiara Caprini, Rob Crittenden, Peter Dunsby,
Ruth Durrer, Lukas Hollenstein, Christos Tsagas and Ethan Siegel
for useful discussions. KT is grateful to Kiyotomo Ichiki
for useful discussion. TK is supported by the JSPS under Contract
No.~01642. The work of TS was supported by Grant-in-Aid for
Scientific Research from Ministry of Education, Science, Sports
and Culture of Japan (No.~13135208, No.~14102004, No.~17740136 and
No.~17340075), the Japan-U.K. and Japan-France Research
Cooperative Program. The work of RM was partly supported by PPARC.
KT is supported by a Grant-in-Aid for JSPS Fellows.




\begin{thebibliography}{99}



\bibitem{Review}
  D.~Grasso and H.~R.~Rubinstein,
  Phys.\ Rept.\  {\bf 348}, 163 (2001)
  [arXiv:astro-ph/0009061];\\
  L.~M.~Widrow,
  Rev.\ Mod.\ Phys.\  {\bf 74}, 775 (2003)
  [arXiv:astro-ph/0207240];\\\
  M.~Giovannini,
  Int.\ J.\ Mod.\ Phys.\ D {\bf 13}, 391 (2004)
  [arXiv:astro-ph/0312614];\\
  M.~Giovannini,
  Class.\ Quant.\ Grav.\  {\bf 23}, R1 (2006)
  [arXiv:astro-ph/0508544];\\
  K.~Subramanian,
  arXiv:astro-ph/0601570;\\
  R.~Durrer,
  arXiv:astro-ph/0609216.

\bibitem{Dilaton}
See for example,
  M.~S.~Turner and L.~M.~Widrow,
  Phys.\ Rev.\ D {\bf 37}, 2743 (1988);\\
  B.~Ratra,
  Astrophys.\ J.\  {\bf 391}, L1 (1992);\\
  A.~Dolgov,
  Phys.\ Rev.\ D {\bf 48}, 2499 (1993)
  [arXiv:hep-ph/9301280];\\
  M.~Gasperini, M.~Giovannini and G.~Veneziano,
  Phys.\ Rev.\ Lett.\  {\bf 75}, 3796 (1995)
  [arXiv:hep-th/9504083];\\
  K.~Bamba and J.~Yokoyama,
  Phys.\ Rev.\ D {\bf 69}, 043507 (2004)
  [arXiv:astro-ph/0310824];\\
  K.~Enqvist, A.~Jokinen and A.~Mazumdar,
  JCAP {\bf 0411}, 001 (2004)
  [arXiv:hep-ph/0404269];\\
  K.~Bamba and J.~Yokoyama,
  Phys.\ Rev.\ D {\bf 70}, 083508 (2004)
  [arXiv:hep-ph/0409237];\\
  K.~Bamba and M.~Sasaki, astro-ph/0611701.

\bibitem{Harrison}
E. R. Harrison, Mon. Not. R. Astron. Soc. {\bf 147}, 279 (1970).

\bibitem{Vachaspati:1991tt}
  T.~Vachaspati and A.~Vilenkin,
  Phys.\ Rev.\ Lett.\  {\bf 67}, 1057 (1991);\\
  D.~N.~Vollick,
  Phys.\ Rev.\ D {\bf 48}, 3585 (1993);\\
  K.~Dimopoulos,
  Phys.\ Rev.\ D {\bf 57}, 4629 (1998)
  [arXiv:hep-ph/9706513];\\
  A.~C.~Davis and K.~Dimopoulos,
  Phys.\ Rev.\ D {\bf 72}, 043517 (2005)
  [arXiv:hep-ph/0505242].

\bibitem{Lewis:2004kg}
  A.~Rebhan,
  Astrophys.\ J.\  {\bf 392}, 385 (1992);\\
  A.~Lewis,
  Phys.\ Rev.\ D {\bf 70}, 043518 (2004)
  [arXiv:astro-ph/0403583].

\bibitem{Riotto}
  S.~Matarrese, S.~Mollerach, A.~Notari and A.~Riotto,
  Phys.\ Rev.\ D {\bf 71}, 043502 (2005)
  [arXiv:astro-ph/0410687].

\bibitem{Gopal:2004ut}
  R.~Gopal and S.~Sethi,
  Mon.\ Not.\ Roy.\ Astron.\ Soc.\  {\bf 363}, 529 (2005)
  [arXiv:astro-ph/0411170].

\bibitem{Takahashi}
  K.~Takahashi, K.~Ichiki, H.~Ohno and H.~Hanayama,
  Phys.\ Rev.\ Lett.\  {\bf 95}, 121301 (2005)
  [arXiv:astro-ph/0502283].

\bibitem{Ichiki}
  K.~Ichiki, K.~Takahashi, H.~Ohno, H.~Hanayama and N.~Sugiyama,
  Science {\bf 311}, 827 (2006)
  [arXiv:astro-ph/0603631].

\bibitem{Fry}
  E.~R.~Siegel and J.~N.~Fry,
  arXiv:astro-ph/0604526.

\bibitem{ITS2007}
  K.~Ichiki, K.~Takahashi, N.~Sugiyama, H.~Hanayama and H.~Ohno,
  arXiv:astro-ph/0701329.

\bibitem{other}
  H.~Lesch and M.~Chiba,
  arXiv:astro-ph/9411072;\\
  C.~J.~Hogan,
  arXiv:astro-ph/0005380;\\
  Z.~Berezhiani and A.~D.~Dolgov,
  Astropart.\ Phys.\  {\bf 21}, 59 (2004)
  [arXiv:astro-ph/0305595];\\
  G.~Betschart, P.~K.~S.~Dunsby and M.~Marklund,
  Class.\ Quant.\ Grav.\  {\bf 21}, 2115 (2004)
  [arXiv:gr-qc/0310085].


\bibitem{Dynamo}
See for example,
  A.~C.~Davis, M.~Lilley and O.~Tornkvist,
  Phys.\ Rev.\ D {\bf 60}, 021301 (1999)
  [arXiv:astro-ph/9904022].

\bibitem{Ellis}
G. F. R. Ellis, in {\em Cargese Lectures in Physics}, vol. VI, ed.
E. Schatzman (Gordon \& Breach, NY, 1973);\\
  C.~G.~Tsagas and J.~D.~Barrow,
  Class.\ Quant.\ Grav.\  {\bf 14}, 2539 (1997)
  [arXiv:gr-qc/9704015];\\
  C.~Caprini, S.~Biller and P.~G.~Ferreira,
  JCAP {\bf 0502}, 006 (2005)
  [arXiv:hep-ph/0310066];\\
  C.~G.~Tsagas,
  Class.\ Quant.\ Grav.\  {\bf 22}, 393 (2005)
  [arXiv:gr-qc/0407080].

\bibitem{Roy98}
  R.~Maartens, T.~Gebbie and G.~F.~R.~Ellis,
  Phys.\ Rev.\ D {\bf 59}, 083506 (1999)
  [arXiv:astro-ph/9808163].

\bibitem{TC}
  P.~J.~E.~Peebles and J.~T.~Yu,
  Astrophys.\ J.\  {\bf 162}, 815 (1970);\\
  P.~J.~E.~Peebles, {\it The Large Scale Structure of the Universe}
  (Princeton University Press, 1980).

\bibitem{Dolgov}
A.~D.~Dolgov and D.~Grasso,
  Phys.\ Rev.\ Lett.\  {\bf 88}, 011301 (2002)
  [arXiv:astro-ph/0106154];\\
  Z.~Berezhiani and A.~D.~Dolgov,
  Astropart.\ Phys.\  {\bf 21}, 59 (2004)
  [arXiv:astro-ph/0305595].


\bibitem{TIS2007}
K.~Takahashi, K.~Ichiki and N.~Sugiyama, in preparation.

\end{thebibliography}
\end{document}